\documentclass[nofootinbib,notitlepage,superscriptaddress,10pt,aps,prd,onecolumn]{revtex4-1}
\usepackage[utf8]{inputenc}
\usepackage{amsmath,mathtools}

\usepackage{tensor}
\usepackage{amssymb}
\usepackage{amsmath}
\usepackage{cases}

\usepackage{graphicx}
\usepackage{physics}
\usepackage{hyperref}
\usepackage{natbib}
\usepackage{lipsum}  
\usepackage{bigints}
\usepackage{verbatim}
\newcommand{\leri}[1]{\left(#1\right)}

\begin{document}
\title{Symmetries of the Electromagnetic Turbulence in a Tokamak
Edge}
\author{Giovanni Montani}
\email{giovanni.montani@enea.it}
\affiliation{ENEA, Fusion and Nuclear Safety Department, C. R. Frascati,
	Via E. Fermi 45, 00044 Frascati (Roma), Italy}
 \affiliation{Physics Department, ``Sapienza'' University of Rome, P.le Aldo Moro 5, 00185 (Roma), Italy}
\author{Fabio Moretti}
\email{fabio.moretti.1@enea.it}
\affiliation{ENEA, Fusion and Nuclear Safety Department, C. R. Frascati,
	Via E. Fermi 45, 00044 Frascati (Roma), Italy}

\begin{abstract}
{We construct the low-frequency formulation of the turbulence characterizing the plasma in a Tokamak edge.
Under rather natural assumptions we demonstrate that, even in the presence of poloidal magnetic fluctuations, 
it is possible to deal with a reduced model for the turbulence dynamics.
This model relies on a single equation for the electric potential, from which all the physical turbulent properties can be calculated. 
The main result of the present analysis concerns the existence of a specific Fourier branch for the dynamics which demonstrate the attractive character of the two-dimensional turbulence with respect to non-axisymmetric fluctuations. 
The peculiar nature of this instability, affecting the non-axially symmetric modes, 
is discussed in some detail, by recovering two different physical regimes.}
\end{abstract}
	\maketitle
\section{Introduction}

One of the most important phenomena affecting the transport processes in a Tokamak edge is the turbulence affecting the plasma in that region \cite{wesson2011tokamaks,tokam3x,stegmeir-GRILLIX-19,2022PhPl...29c2302C}. This turbulent behavior is due to a number of different physical ingredients \cite{Oliveira_2022,Graves_2005,Zweben_2007}, but one of the most relevant contributions is certainly due to the 
so-called non-linear drift response, as discussed in \cite{scott02}, see also the review \cite{Scott_2007}. 
The drift turbulence is a non-linear self-sustained process, which receive a linear triggering in terms of free energy available to the spectrum development from the background pressure gradient \cite{hase-waka83,hase-waka87,hase-mima18,Diamond_2011}. 
However, the non-linear dynamics of the system is then independent of this linear triggering, even because this source of linear instability is significantly suppressed in the presence of magnetic shear \cite{scott02}. 

The theoretical ground of the non-linear drift response is a two-fluid model 
(electrons and ions), which is justified by the plasma conditions (low density and temperature) of the edge plasma and the assumption of a low-frequency physics, in which the typical time scale of the turbulent processes is long in comparison to the inverse of the ion gyrofrequency. The original Hasegawa-Wakatani formulation describing the drift turbulence dynamics in \cite{hase-waka83,hase-waka87} has been based on the effective representation of the Fourier evolution, de facto reducing the problem to a two-dimensional picture \cite{R_H_Kraichnan_1980,annurev:/content/journals/10.1146/annurev-fluid-120710-101240,10.1063/1.1762301,Kraichnan_1975}. The limits of this representation and the impact of dealing with a realistic three-dimensional scenario have been discussed in \cite{biskamp95} (for a recent systematic 
analysis on this aspect see \cite{2024arXiv240509837C}).

However, we remark that despite the simple electrostatic formulation of the turbulence captures many real features of the non-linear dynamics in currently operating systems nonetheless, as the 
plasma $\beta$-parameter increases enough, a poloidal fluctuation of the magnetic field becomes a relevant subject of the turbulent behavior \cite{scott02}. 

In a recent work \cite{2023PhyD..45133774M}, see also \cite{2022Fluid...7..157M}, it has been proposed an electrostatic reduced model, which leads to a single equation for the electrostatic potential field, as a result of neglecting the linear instability trigger and 
taking a unique coefficient to describe both the shear viscosity and the particle diffusion, see \cite{biskamp95}. 
As important issue, this reduced model was able to predict the attractive character of the two-dimensional 
axisymmetric configuration with respect to the parallel fluctuations. 
Since the magnetic field has been 
taken constant and uniform along the axial direction, the obtained picture can be qualitatively associated to the spectral features of the plasma close enough to the X-Point of the magnetic configuration of a Tokamak device.

Here, starting with a rather general picture of the low-frequency turbulence, we then implement a number of motivated simplifications which again lead to a reduced model, characterized by a single equation for the electric field potential. The main difference of the present model with respect to previous ones, having the same scheme, is that here we include the axial potential vector amid fluctuations, which is found to be responsible for a turbulent poloidal magnetic field. Like in previous studies \cite{2023Symm...15.1745C,2023PhyD..45133774M}, a constitutive relation links the number density fluctuation to the vorticity field. Moreover, we 
show that it is possible to evaluate the axial vector potential from the electric field, via one of the basic dynamical equations. 

We provide an analytical study, showing the attractivity of the two-dimensional axisymmetric spectrum also in the presence of a turbulent poloidal magnetic field. In fact, we determine a decaying branch of the non-axially symmetric perturbation, as an exact solution of the considered dynamics in the Fourier space. 
We observe that the 
obtained decaying behavior has two different characteristic regions of the 
dispersion relation, associated to a non-zero and zero real frequency, respectively. This two decaying regimes both differ from the corresponding spectral behavior of the pure electrostatic case and, overall, they are not directly connected to the value of the toroidal wave number $n$: below a given n-dependent 
wave number the decaying rate is not affected at all by $n$ itself, while a dependence emerges in toroidal spectral region above this value (anyway the dependence is different from the electrostatic case). 

This result offers an interesting scenario also for incoming large size Tokamak machines, like Italian DTT \cite{dtt19,DTT_2021} or ITER. In fact, we can conclude that, close enough the X-point configuration, the turbulence has mainly a two-dimensional spectral feature even when the 
$\beta$-parameter of the plasma is 
considerably large with respect to 
currently operating Tokamak \cite{Oliveira_2022,Graves_2005}. 
The manuscript is organized as follows: in Section \ref{sec2} we present the hypotheses on which our model of electromagnetic edge turbulence is grounded, deriving the basic dynamical equations and formulating the problem in adimensional variables; in Section \ref{sec3} we recast the set of PDE obtained in order to highlight the fact that it represents a generalized version of the Hasegawa-Wakatani model; in Section \ref{sec4} we make further assumptions, well justified in the case of a Tokamak plasma scenario, showing how the system of PDE results condensed in a single governing equation for the electric potential. Fourier analysis is then performed on this unique equation, outlining the presence of two distinct branches of the dispersion relation. Conclusions are finally drawn in Section \ref{sec5}. 

\section{Construction of the model}
\label{sec2}
We analyze the dynamical and physical 
properties of the plasma laying in the edge region of a Tokamak machine with particular reference to 
the magnetic X-point zone, placed out of the separatrix.
We consider a quasi-neutral hydrogen-like 
plasma, in the low-frequency limit
(i.e., the phenomenon evolution rates are much smaller 
than the ion gyrofrequency $\Omega_i$) 
and we provide a two-fluid representation 
of its dynamics, within the so-called "drift ordering approximation''. 

The equilibrium magnetic configuration 
is here modeled via a constant and 
uniform magnetic filed. Choosing a set 
of Cartesian coordinates $\{ x,y,z\}$ 
(the relative versors are denoted by 
$\textbf{e}_x$, $\textbf{e}_y$ and 
$\textbf{e}_z$, respectively), the 
background magnetic field is expressed as $\textbf{B} = B_0\textbf{e}_z$, 
with $B_0 = const.$ (henceforth, the 
suffix $0$ will denote background 
quantities). The $z$-axis must be thought as 
the toroidal axis of a Tokamak device.

In the proposed approximation scheme 
and neglecting diamagnetic effects, 
the electron momentum balance 
provides for the orthogonal 
electron velocity $\textbf{v}_{\perp}$ 
the following $\mathbf{E} \times \mathbf{B}$ expression: 
\begin{equation}
	\textbf{v}_{\perp}\equiv 
	\textbf{v}_E = \frac{c}{B_0}\left( -\partial_y\phi\textbf{e}_x 
	+ \partial_x\phi\textbf{e}_y\right)
	\, , 
	\label{gm1}
\end{equation}
where $\phi (t,x,y,z)$ denotes the electric potential fluctuation (here we neglect any background contribution).
In what follows, the advective (Lagrangian) derivative is constructed with 
$\textbf{v}_E$, namely

\begin{equation}
	\frac{df}{dt} \equiv 
	\partial_tf + \frac{c}{B_0}\left(\partial_x\phi\partial_yf - \partial_y\phi\partial_xf\right)
	\, , 
	\label{gm2}
\end{equation}
in which $f$ is a generic scalar or vector function.
The ion momentum balance provides the 
dynamical equation for the ion perpendicular velocity $\textbf{u}_{\perp}$

\begin{equation}
	\frac{d\textbf{u}_{\perp}}{dt} = \frac{e}{M_i}\left( - \nabla_{\perp}\phi + \frac{\textbf{u}_{\perp}}{c}\wedge \textbf{e}_z B_0\right) 
	+ \nu\Delta_{\perp}\textbf{u}_{\perp}
	\, , 
	\label{gn3}
\end{equation}
where $e$ is the elementary charge, $M_i$ the ion mass and $\nu$ denotes the 
constant kinematic (perpendicular) ion viscosity.

If, in agreement with the low frequency approximation, we set $\textbf{u}_{\perp} = 
\textbf{v}_E + \textbf{u}^{(1)}_{\perp}$, where the correction to the $\mathbf{E} \times \mathbf{B}$-velocity is taken small, then Eq. 
(\ref{gn3}) gives, at first order,
\begin{equation}
	\textbf{u}^{(1)}_{\perp} = 
	\textbf{e}_z\wedge 
	\frac{1}{\Omega_i}\left( \frac{d\textbf{v}_E}{dt} - \nu \Delta_{\perp}\textbf{v}_E\right)
	\, , 
	\label{gm3}
\end{equation}
in which the first term in parenthesis 
corresponds to the so-called polarization drift velocity. 

The charge conservation equation 
$\nabla _{\perp}\cdot \textbf{j}_{\perp} = - \nabla_{\parallel}\cdot \textbf{j}_{\parallel}$ can be easily 
restated by observing that 
$\textbf{j}_{\perp} = n_0e\textbf{u}^{(1)}_{\perp}$, where we indicate with $n_0$ the electron and ion background density. Hence, we get the following 
equation:

\begin{equation}
	\partial_t\Delta_{\perp}\phi + 
	\frac{c}{B_0}\left(\partial_x\phi \partial_y\Delta_{\perp}\phi - \partial_y\phi\partial_x\Delta_{\perp}\phi\right)  
	= 4\pi\frac{v_A^2}{c^2}\partial_zj_z 
	+ \nu\Delta^2_{\perp}\phi 
	\, , 
	\label{gm5}
\end{equation}
where $v_A$ denotes the background Alfvén velocity and we expressed the parallel gradient as $\partial_z$, 
while $\Delta_{\perp}$ stands for $\partial_x^2 + \partial_y^2$. 
It is worth noting that the electric field dynamics is not affected by 
the electron and ion diamagnetic velocities which, for a constant magnetic 
field, turn out to be divergenceless. 

In what follows, we assume a background 
pressure profile depending on the coordinate $x$ only and such that 
$dp_0/dx = -p_0/l_0$, where $l_0$ is a fixed 
spatial length. By other words, we 
consider an exponential behavior of the 
background pressure, which would require, 
for a precise force balance, an additional 
dependence of the magnetic field $z$-component of the form 
$4\pi p_0/B_0$, which will be neglected 
in what follows since in a Tokamak 
the plasma $\beta$-parameter is rather small. 
If we denote by $\bar{p}$ the pressure 
contrast (fluctuation value to the 
background one), it obeys the following 
equation (the same for ions and electrons):

\begin{equation}
	\partial_t\bar{p} + \frac{c}{B_0}\left( \partial_x\phi\partial_y\bar{p} - \partial_y\phi\partial_x\bar{p}\right) = -\frac{c}{B_0l_0}\partial_y\phi +
	\partial_z\left(\frac{1}{n_0 e}j_z - u_z\right)\, , 
	\label{gm6}
\end{equation} 
being $u_z$ the ion parallel fluctuating velocity. 
The equation governing the dynamics of $u_z$ 
reads 

\begin{equation}
	\partial_tu_z + \frac{c}{B_0}\left(\partial_x\phi\partial_yu_z - 
	\partial_y\phi\partial_xu_z \right) = -  
	\frac{2K_BT_0}{M_i}\partial_z\bar{p}
	\, . 
	\label{gm7}
\end{equation}
where $T_0$ is the background temperature common to ions and electrons, linked to the background pressure via $p_0=n_0 K_B T_0$, with $K_B$ the Boltzmann constant.
If we denote the perturbation of the parallel 
magnetic vector potential as $\textbf{A}_{\parallel} 
\equiv \psi \textbf{e}_z$, then we 
easily get from the Ampere law (displacement  
currents are negligible in a non-relativistic plasma) 
the relation

\begin{equation}
	\textbf{j}_{\parallel} = -\frac{c}{4\pi}\Delta_{\perp}\psi 
	\, \textbf{e}_z \, ,
	\label{gm8}
\end{equation}
connecting the parallel current to the turbulent fluctuating magnetic field.
The system describing the perturbations dynamics 
is completed by the generalized Ohm law, 
taking the explicit form
\begin{equation}
	\frac{1}{c}\partial_t\psi = 
	\frac{c}{4\pi\sigma}\Delta_{\perp}\psi 
	+ \partial_z\left( \frac{K_BT_0}{e}\bar{p} - \phi\right)
	\, , 
	\label{gm9}
\end{equation}
where $\sigma\equiv 1.96\, n_0 e^2/m_e \nu_{ei}$ denotes the electric parallel 
conductivity, with $m_e$ the electron mass and $\nu_{ei}$ the electron-ion collision frequency. 
We conclude the construction of this 
dynamical scenario by stressing the 
relations between the electric field components and the potentials, i.e. 
$\textbf{E}_{\perp} = -\nabla_{\perp}\phi$ and $E_z = 
-\partial_z\phi - \frac{1}{c}\partial_t\psi$.

We now aim to formulate the dynamical 
system, derived above, by means of 
dimensionless quantities. To this end, we introduce new space and time coordinates as $u\equiv x/L$, $v\equiv y/L$, $w\equiv (2\pi z)/R$ and 
$\tau\equiv \Omega_it$. Here $L$ and $R$ are two spatial scales (we assume $R \gg L$) characterizing the poloidal plane and the toroidal direction respectively.
Analogously, we define the new set of dynamical variables as $\Phi \equiv e\phi /K_BT$, 
$\Psi \equiv e\psi/K_BT$ and 
$U\equiv (2\pi u_z)/(\Omega_iR)$. 
The fourth unknown $\bar{p}$ is dimensionless by definition.
We also introduce $D_{\perp}\equiv \partial_u^2 + \partial_v^2$.

In terms of this set of redefined coordinates and variables, Eqs.(\ref{gm5}), 
(\ref{gm6}), (\ref{gm7}) and (\ref{gm9}) take the explicit form:
\begin{align}
&\partial_{\tau}D_{\perp}\Phi + A\left\{\Phi , D_{\perp}\Phi\right\}
	= - BD_{\perp}\partial_w\Psi + CD_{\perp}^2\Phi
	\, , \label{10a}\\
 &\partial_{\tau}\bar{p} + 
	A\left\{\Phi,\bar{p}\right\} =-r_1A\partial_v\Phi -
	\partial_w\left(DD_{\perp}\Psi + U\right)
	\, , \label{11a} \\
 &\partial_{\tau}U + A\left\{ 
	\Phi ,U\right\} = - 2r_2A\partial_w\bar{p} \, , 
	\label{12} \\
 &\partial_{\tau}\Psi = 
	ED_{\perp}\Psi + F\partial_w\left( \bar{p} - \Phi\right)
	\, , \label{13a}
\end{align}
where we introduced the dimensionless version of the constants characterizing the physical features of our model as
\begin{align}
    &A\equiv \frac{K_BT_0}{M_i\Omega_i^2L^2}\, , \quad B\equiv \frac{2\pi v_A^2}{c\Omega_iR}\, , \quad C\equiv \frac{\nu}{\Omega_iL^2}
	\, , \\
 &D\equiv \frac{2 \pi c\lambda_D^2}{L^2\Omega_iR}
	\, ,\quad E\equiv \frac{c^2}{4\pi\Omega_i\sigma L^2}\, ,\quad F\equiv \frac{2 \pi c}{\Omega_iR}
	\, ,\\
 &\qquad \qquad \quad r_1\equiv \frac{L}{l_0}\, ,\quad r_2\equiv \frac{4 \pi ^2 L^2}{R^2}. 
\end{align}
The standard symbol $\lambda_D$ indicates the Debye length and 
we observe that $A\equiv \bar{\rho}_i^2$, 
being $\bar{\rho}_i$ the ion Larmor radius normalized with the poloidal length scale, i.e. $\bar{\rho}_i=\rho_i/L$.
Finally, we also 
adopted the compact notation

\begin{equation}
	\left\{ \Phi ,f\right\} 
	\equiv \partial_u\Phi\partial_v f - \partial_v\Phi\partial_u f 
	\, ,
	\label{gm14}
\end{equation}
where $f$ is a generic scalar function of the spatial coordinates $(u,v)$.
The system above corresponds to a closed set of partial differential equations 
in the four unknowns $\Phi$, $\bar{p}$, $U$ and $\Psi$, which, once assigned suitable initial and boundary conditions (see below),  provides a description of the plasma turbulent dynamics. 

\section{Generalized Hasegawa-Wakatani model}\label{sec3}

In this section we show that the set of dynamical equations derived above is equivalent to a generalized  Hasegawa-Wakatani model. Indeed, by expressing the 
quantity $D_{\perp}\Psi$, i.e. the 
current density, from Eq. (\ref{13a}) 
into Eqs (\ref{10a}) and (\ref{11a}), we get the following two restated 
equations
\begin{align}
	&\partial_{\tau}D_{\perp}\Phi + \rho_i^2\left\{\Phi , D_{\perp}\Phi\right\}
	= -\iota \partial_{\tau}\partial_w\Psi + \eta \partial_w^2\left( 
	\bar{p} - \Phi\right) + CD_{\perp}^2\Phi
	\, , \label{gf1a}\\
 &\partial_{\tau}\bar{p} + 
	\rho_i^2\left\{\Phi,\bar{p}\right\} = -r_1\rho_i^2\partial_v\Phi -
	\delta \partial_{\tau}\partial_w \Psi + \epsilon \partial^2_w 
	\left(\bar{p} - \Phi\right) - 
	\partial_w U + \mathcal{D}_pD_{\perp}\bar{p}
	\, , \label{gf2a}
\end{align}
 where we redefined the coupling constants of our model as
\begin{align}
	&\iota \equiv \frac{B}{E}\, ,\; 
	\eta \equiv \frac{BF}{E},
	\, 
	\label{gf1b}\\
	&\delta \equiv \frac{D}{E} 
	\, ,\; 
	\epsilon \equiv \frac{DF}{E}
	\, .
	\label{gf2b}
\end{align}
In Eq. (\ref{gf2a}), we added a diffusion term ($\mathcal{D}_p$ being a constant diffusion coefficient) to account for different transport regimes in a 
Tokamak edge plasma.

It is worth noting that, in our scheme, the kinetic viscosity $\nu$ and the 
electron-ion collision frequency 
$\nu_{ie}$ are constants, independent of the spatial point and instant of time considered. Hence the parameters 
$\iota$, $\eta$, $\delta$ and $\epsilon$ are mere positive real numbers and the 
set (\ref{12}), 
(\ref{gf1a}) and (\ref{gf2a}) 
is a system of PDE with constant 
coefficients. The same does not 
hold for Eq. (\ref{13a}): indeed, in this case it is present an explicit dependence on the specific point considered, due to the background density $n_0(x)$ appearing in the definition of the parallel conductivity $\sigma$. In order to deal with a system of PDE with constant coefficients one can approximate the number density $n_0(x) = n^*e^{-x/l_0}$ ($n^*$ denoting the 
X-point number density) with its mean value on the interval $-l_0/2 \le x \le l_0/2$, resulting in $n_0=2n^*\sinh \left( 1/2 \right )$.

Now, as fourth equation for the four unknowns $\Phi$, $\bar{p}$, $U$ and $\Psi$, instead of Eq. (\ref{13a}), we 
consider Eq. (\ref{10a}) which we recast as
\begin{equation}
	D_{\perp}\left( \partial_{\tau}\Phi + B\partial_w\Psi\right) = -\rho_i^2\left\{ \Phi \, ,\, D_{\perp}\phi \right\} + C D^2_{\perp}\Phi
	\, .
	\label{gfy1}
\end{equation}

We observe that, in the linearized inviscid case, the equation above would correspond to the Lorentz gauge condition, as restricted to the present symmetry.

\section{Relevant reduced model}
\label{sec4}

We now focus the attention to 
a reduced model, described 
by a single equation for the electric potential. To this end we 
neglect, in the system (\ref{gf1a}), (\ref{gf2a}) and (\ref{12}), two quantities that 
are usually small in many turbulence 
configurations of a Tokamak, i.e. 
the gradient of the background pressure $p_0$ (responsible for a linear 
instability) and the presence of 
the velocity $U$ in the problem. Hence, we take the limit $l_0\rightarrow \infty$ in Eqs. (\ref{gf1a})-(\ref{gf2a}) and we neglect equation (\ref{12}). Under these assumptions, Eq. (\ref{gf2a}) rewrites as 
\begin{equation}
	\partial_{\tau}\bar{p} 
	+ \rho_i^2\left\{ \Phi ,\bar{p}\right\} = -\delta \partial_{\tau}\partial_w\Psi + \epsilon \partial^2_w\left(\bar{p} - \Phi\right) + \mathcal{D}_pD_{\perp}\bar{p}
	\, . 
	\label{feq4}
\end{equation}
If we now set $\mathcal{D}_p \equiv C$ we notice that Eqs. (\ref{gf1a}) and (\ref{feq4}) result identical, provided the constitutive relation
\begin{equation}
	D_{\perp}\Phi = \frac{B}{D}\bar{p}
	\, 
	\label{feq5}
\end{equation}
is implemented. Inserting (\ref{feq5}) into the equation for the electric potential (\ref{gf1a}) yields
\begin{equation}
	\partial_{\tau}D_{\perp}\Phi + \rho_i^2\left\{ \Phi \, ,D_{\perp}\Phi \right\} 
	= - \iota \partial_{\tau}\partial_w\Psi + \eta \partial^2_w\left( q D_{\perp}\Phi - \Phi \right) 
	+ CD^2_{\perp}\Phi 
	\, ,
	\label{gfy2}
\end{equation}
with $q\equiv D/B$. In this way we reduced the dynamical system composed by Eqs. (\ref{12}), (\ref{gf1a}), (\ref{gf2a}) and (\ref{gfy1}) to a set of  two coupled equations (\ref{gfy1}) and (\ref{gfy2}), equipped with the supplemental constitutive relation (\ref{feq5}). Now, if we apply to Eq. (\ref{gfy2}) the operator $D_{\perp}$ and we make use of Eq. (\ref{gfy1}), we obtain a single equation characterizing our model, in which the only unknown is the electric potential, namely 
\begin{equation}
     \partial_{\tau}\left( D_{\perp}-\frac{\iota}{B}\partial_{\tau}\right) D_{\perp} \Phi + \rho_i^2\left( D_{\perp} - \frac{\iota}{B}\partial_{\tau}\right)\left\{ \Phi \, ,D_{\perp}\Phi \right\}
 =\eta \partial^2_wD_{\perp}\left( q D_{\perp}\Phi - \Phi \right) 
	+ C\left( D_{\perp}-\frac{\iota}{B}\partial_{\tau}\right) D^2_{\perp}\Phi 
	\, .
	\label{gfy3}
\end{equation}
We remark that the terms times $\iota /B$ vanish for the axisymmetric component of the theory, for the fact that they represent derivatives along the toroidal direction of the vector potential. 

We are now interested in analyzing the properties of equation (\ref{gfy3}). To this end, we exploit the periodicity of the system on the coordinate $w$ and we express the electric potential through a Fourier series along the toroidal direction. The poloidal plane coordinates $(u,v)$ will be treated instead with a Fourier transform, in order to obtain 

\begin{equation}
	\Phi (\tau , u,v,w) = \sum _{n=-\infty}^{\infty}\int_D \frac{dk_udk_v}{(2\pi )^2} \,\xi^n_{\mathbf{k}} \, e^{i(k_uu+k_vv+nw)}
	\, ,
	\label{gfy4}
\end{equation}
where $\xi^n_{\mathbf{k}} = \xi^n (\tau , k_u,k_v)$ and $D$ is a circular domain of radius $k_{max} = 2\pi /\rho_i$, corresponding to the cut-off value of the drift-fluid approximation here addressed.

Substituting into Eq. (\ref{gfy3}), we get the following basic 
equation for the Fourier modes
\begin{multline}
	k^2\left( k^2 + p\partial_{\tau}\right) 
	\partial_{\tau}\xi_{\mathbf{k}}^n + 
	\rho_i^2\left( k^2 + p \partial_{\tau}\right) 
	\sum _{m=-\infty}^{\infty}\int \frac{dq_udq_v}{(2\pi )^2} 
	q^2\left( k_uq_v - k_vq_u\right) 
	\xi_{\mathbf{k}-\mathbf{q}}^{n-m}\xi_{\mathbf{q}}^m = 
	 \\
	=- \eta n^2k^2\left( qk^2 + 1\right) \xi_{\mathbf{k}}^n 
	- Ck^4\left( k^2 + p \partial_{\tau}\right) \xi_{\mathbf{k}}^n 
	\, , 
	\label{gf5}
\end{multline}
where $p\equiv \iota /B$ and $k=\sqrt{k_u^2+k_v^2}$ is the norm of the poloidal wavenumber. It is easy to verify that the $n=0$ mode satisfies 
\begin{equation}
k^2\partial_{\tau}\xi_{\mathbf{k}}^0 + \rho_i^2 
	\int_D\frac{dq_udq_v}{(2\pi )^3}q^2
	\left(k_uq_v-k_vq_u\right) \xi^0_{\mathbf{k}-\mathbf{q}}\xi^0_{\mathbf{q}} + Ck^4\xi^0_{\mathbf{k}} = 0
	\, .
	\label{gfye1}
\end{equation}
In the inviscid limit ($C\equiv 0$), the equation above 
is characterized by a statistical equilibrium \cite{seyler75,2022Fluid...7..157M}, 
corresponding to an energy spectrum of the form
\begin{equation}
	E(k) = \frac{1}{\alpha + \beta k^2}
	\, , 
	\label{ggfy1}
\end{equation}
where the parameters $\alpha$ and $\beta$ represent two inverse temperatures, associated to 
the energy and enstrophy constants of motion, respectively.

As in \cite{2022Fluid...7..157M}, we now concentrate our attention to 
the case in which $\alpha$ can be neglected in the 
energy spectrum (\ref{ggfy1}) (this choice is linked to 
a specific structure of the system initialization), 
for which we deal with a constant enstrophy isotropic spectrum. The corresponding mode amplitude results in
\begin{equation}
	\xi ^0_k = \frac{\Gamma}{k^2}
	\, , 
	\label{ggfy2}
\end{equation}
where $\Gamma$ is a complex constant. It  is easy to 
verify that the field form (\ref{ggfy2}) is a steady solution of Eq. (\ref{gfye1}) in the inviscid case $C\equiv 0$ (for details see the Appendix). 
Clearly, it remains a good approximate 
solution also for the viscous case, when $Ck^2\mid \Gamma\mid \ll 1$, and this 
regime holds valid for a significant part of the inertial region spectrum.

We now linearize Eq. (\ref{gf5}) in 
the generic $n\neq 0$ mode, when considering the steady solution 
(\ref{ggfy2}) as the dominant contribution, namely
\begin{eqnarray}
	k^2\left( k^2 + p\partial_{\tau}\right) 
	\partial_{\tau}\xi_{\mathbf{k}}^n + 
	\rho_i^2\left( k^2 + p \partial_{\tau}\right) 
	\int \frac{dq_udq_v}{(2\pi )^2}  
	q^2\left( k_uq_v - k_vq_u\right) 
	\left[ \xi_{\mathbf{k}-\mathbf{q}}^n
	 \xi_{\mathbf{q}}^0 + \xi_{\mathbf{k}-\mathbf{q}}^0\xi_{\mathbf{q}}^n \right] = 
 	\nonumber \\
	-\eta n^2 k^2\left( qk^2 + 1\right) \xi_{\mathbf{k}}^n 
	- Ck^4\left( k^2 + p \partial_{\tau}\right) \xi_{\mathbf{k}}^n 
	\, . 
	\label{gf5b}
\end{eqnarray}
By assuming isotropy in the poloidal plane, i.e. $\xi_{\mathbf{k}}^n = f^n(k,\tau )$, we see that the term containing the integral identically vanishes (details in Appendix) and the dynamical equation for the Fourier modes $f^n$ results in

\begin{equation}
	\left( k^2 + p \partial_{\tau}\right) \partial_{\tau}f^n = 
	- \eta n^2\left( qk^2 + 1\right)f^n - Ck^2\left( k^2 + p \partial_{\tau}\right)f^n
	\, .
	\label{ggfy3}
\end{equation}

We are interested in analyzing the dispersion relation and the damping rate for these modes, hence we search for a solution of the above
equation in the form 
$f^n = \bar{f}^n \exp \{ - i \Omega (k)\tau \}$ ($\bar{f}^n$ being a constant and $\Omega$ a complex number), obtaining

\begin{equation}
	p\Omega^2 + (1 + Cp)k^2i\Omega 
	-\eta n^2( qk^2 + 1) - Ck^4 = 0
	\, .
	\label{dr1}
\end{equation}
We explicitly denote the real and imaginary part of $\Omega$ as $\omega$ and $\gamma$ respectively, i.e.
$\Omega = \omega + i \gamma$ 
($\omega$ and $\gamma$ being real functions of $k$). Then, equation (\ref{dr1}) splits into the following system

\begin{figure}
\centering
\includegraphics[width=8.5cm]{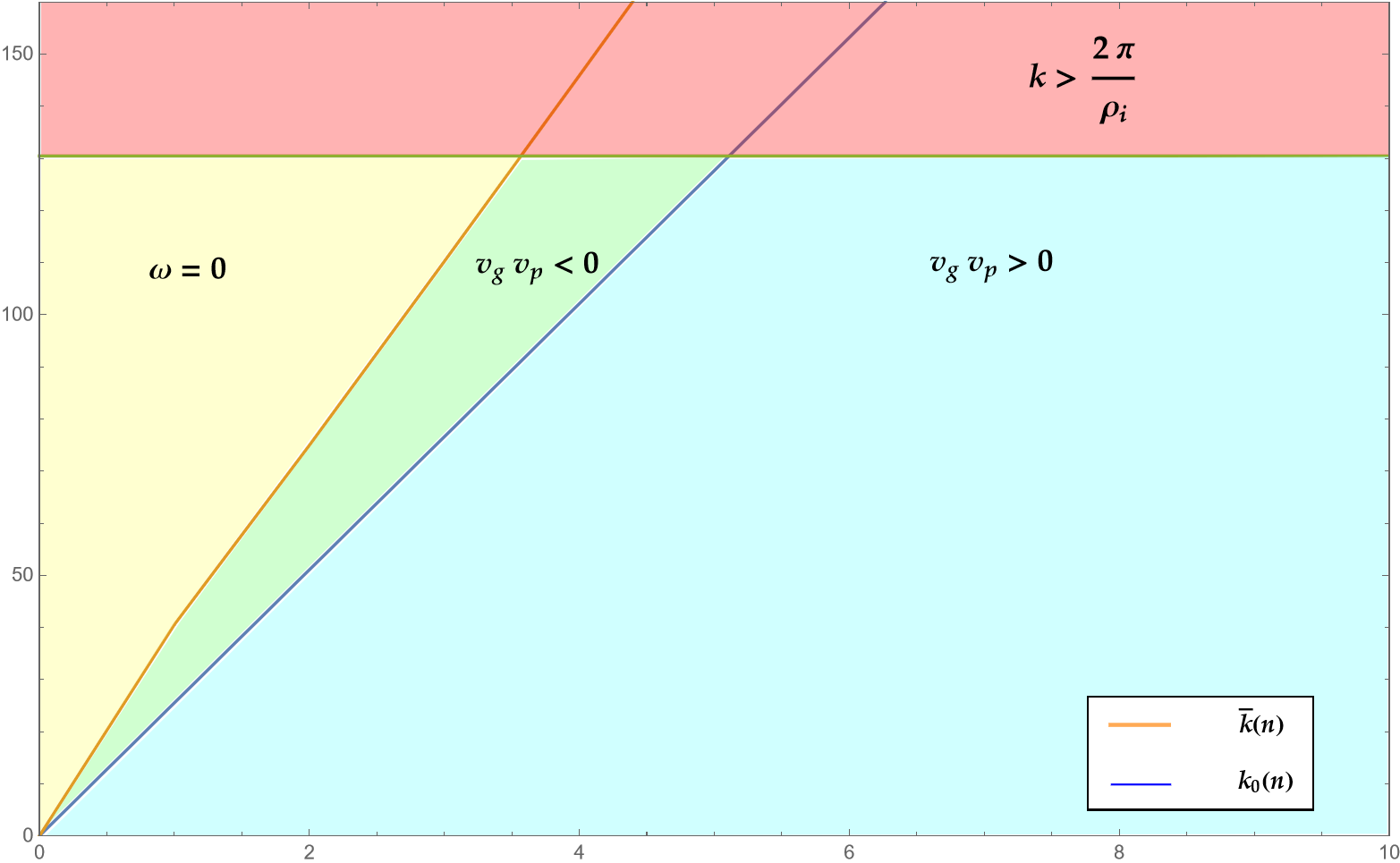}
\caption{Different regimes for the solution of (\ref{dr1}): the area in yellow indicates the solution with $\omega=0$, the sectors in green and light blue represent wave-like solutions having group and phase velocity discordant or concordant, respectively. The orange and blue lines depict the characteristic wavenumbers $\bar{k}_n$ and $k^0_n$ as functions of the toroidal wavenumber $n$, the latter ranging on the horizontal axis. On the vertical axis we plot the norm of the poloidal wavenumber $k=\sqrt{k_u^2+k_v^2}$. The region in red represents values of $k$ greater than the cut-off $\, 2\pi/\rho_i$, for which the fluid description adopted in this work can not be enforced.}
\label{grafvgvp}
\end{figure}

\begin{numcases}{}
   \label{redisp} p\leri{\omega^2-\gamma^2}-\leri{1+C p}k^2 \gamma -\eta n^2
\leri{qk^2+1}-Ck^4=0 \\
\label{imdisp}
\omega \leri{2p\gamma+\leri{1+Cp}k^2}=0.
\end{numcases}
It is immediate to notice that \eqref{imdisp} is satisfied either for $\omega=0$ or $\gamma=-\frac{1+Cp}{2p}k^2$. Inserting the $\omega=0$ solution in \eqref{redisp} yields 
\begin{equation}
    \gamma_{\pm}=-\frac{1+Cp}{2p}k^2 \pm \frac{1}{2p}\sqrt{\leri{1-Cp}^2k^4-4\eta p n^2 \leri{qk^2+1}}
\end{equation}
and the reality of $\gamma_\pm$ is ensured by
\begin{equation}\label{condreal}
    k \geq \bar{k}_n=\sqrt{\frac{2\eta n^2 p q}{\leri{1-Cp}^2}\leri{1+\sqrt{1+\frac{\leri{1-Cp}^2}{\eta n^2 p q^2}}}}.
\end{equation}
We remark that, when \eqref{condreal} is satisfied, both $\gamma_\pm$ are real negative numbers and the corresponding solution describes damped oscillations rather than wave-like fluctuations. 
Conversely, by selecting the second solution of \eqref{imdisp}, i.e. $\gamma=-\frac{1+Cp}{2p}k^2$, and solving \eqref{redisp} for the angular frequency $\omega$, we obtain

\begin{equation}
        \omega_\pm=\pm \frac{1}{2p}\sqrt{-\leri{1-Cp}^2 k^4+4\eta p n^2 \leri{qk^2+1}},
\end{equation}
and in this case, the reality condition on $\omega_\pm$ is satisfied for $k < \bar{k}_n$. It is interesting to compute the group velocity for these wave-like solutions, which reads 
\begin{equation}
    v_g=\mp \frac{k}{p}\frac{\leri{1-Cp}^2k^2-2\eta n^2 pq}{\sqrt{-\leri{1-Cp}^2k^4+4\eta p n^2 \leri{qk^2+1}}}.
\end{equation}
We outline that the group velocity has opposite sign with respect to the phase velocity $v_p=\frac{\omega}{k}$ when 
\begin{equation}
    k>k^0_n=\sqrt{\frac{2\eta n^2 p q }{\leri{1-Cp}^2}},
\end{equation}
whereas both velocities share the same sign in the opposite case $k<k^0_n$. Lastly, for $k=k^0_n$, we deal with a wave-like solution characterized by a null group velocity, i.e. no energy transport is allowed in this regime. It must be stressed that $k^0_n$ is always smaller than $\bar{k}_n$, therefore, for any given $n>0$, one can always select a wavenumber $k$ such that the wave-like solutions associated to $\omega_\pm$ have concordant, discordant or null group velocity. In order to provide some quantitative examples of the phenomenology here described, let us assume typical values of a Tokamak-like environment \cite{2024arXiv240509837C,dtt19,DTT_2021} for the physical parameters characterizing our model: specifically, we set $T=100$ eV, $B_0=3$ T and $n_0=5\times10^{19}$ m$^{-3}$ \cite{dtt19}, thus $\Omega_i\simeq1.4\times 10^8$ s$^{-1}$ and $\rho_i\simeq0.048$ cm. Moreover, we select for the length scales $L$ and $R$ values equal to $1$ cm and $1300$ cm, respectively. We report, in Fig. \ref{grafvgvp}, the different regimes for the solutions of (\ref{dr1}) described above, highlighting the fact that either a pure damped oscillatory fluctuation or wave-like decaying propagating ripples are present, depending on the specific choice of the toroidal an poloidal wavenumbers considered. For instance, the axisymmetric $n=0$ solution admits solely a pure damped oscillatory regime; for $1 \leq n \leq 3$ all three regimes (either oscillatory or wave-like with concordant or discordant velocities) are feasible and the specific configuration depends on the poloidal wavenumber $k$; for $n$ equal to $4$ and $5$ only the wave-like solutions are allowed, whereas for $n>5$ the system has a unique solution, corresponding to the wave-like case with concordant velocities.

\begin{figure}
\centering
\includegraphics[width=8.5cm]{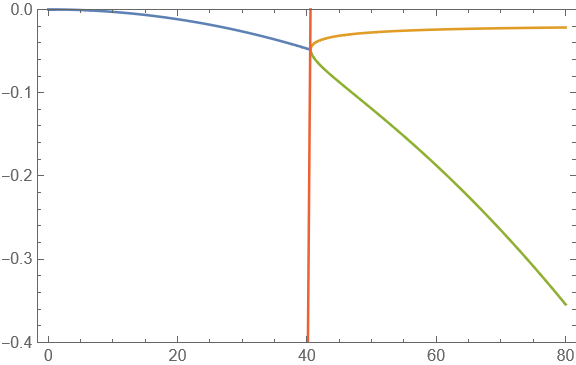}
\caption{Damping coefficients for a value of the toroidal wavenumber $n=1$: the blue curve is relative to the wave-like regime ($\omega \neq 0$), whereas the orange and green curves represent $\gamma_+$ and $\gamma_-$, respectively, characterizing the pure damping regime ($\omega=0$). The red vertical line indicates the value of the characteristic wavenumber $\bar{k}_n$ separating the two regimes. We plot the norm of the poloidal wavenumber $k$ on the horizontal axis and the value of the damping coefficient on the vertical one.}
\label{grafgamma}
\end{figure}

 Then, in Fig. \ref{grafgamma}, we show the behavior of the damping coefficients as functions of the poloidal wavenumber $k$, outlining again the transition between wave-like and oscillatory regimes marked by the threshold wavenumber $\bar{k}_n$.

\section{Concluding remarks}
\label{sec5}

We developed a low-frequency formulation for the two.fluid turbulence in the edge region of a Tokamak plasma, whose magnetic configuration was associated to a constant and uniform axial magnetic field, i.e. well-representing a region very close to the X-point \cite{10.1063/1.4935115}. 

Then, under suitable assumptions, based on the standard operational regimes of medium-size and incoming experiments, we reduced to the dynamics to a simplified model: 
actually, we deal with a single equation for the electric potential field, a constitutive relation giving the pressure fluctuations and an evolutionary 
link between the scalar (electric) and 
vector (de facto the axial component) potentials.

In order to better focus on the physical character of these assumptions, we recall that they can be summarized in the following three statements: i) the background pressure gradients are negligible; 
ii) the ion axial velocity can be 
neglected with respect to the corresponding contribution for electrons; 
iii) the shear viscosity and particle diffusion coefficients are taken comparable. 
The first assumption is justified by the fact that the background pressure gradient, in a realistic Tokamak configuration, is naturally suppressed by 
means of magnetic shear \cite{Scott_2007}. 
The second hypothesis provides a good representation of the Tokamak typical plasmas in the absence of spontaneous rotation \cite{10.1063/1.4867656} or of 
neutral beam injection. Finally, 
the possibility to take the viscosity and diffusivity coefficients of the same order has been considered in various formulations, see for instance \cite{biskamp95}, and this is due to the role that the higher order Laplacian operators take in the edge turbulence picture. 
In fact, the value of these coefficients is relevant in stabilizing the numerical simulation and, de facto, their amplitude can not be \emph{a priori} 
predicted, simply because it is 
directly and significantly influenced by the turbulence features. Thus the 
value of the dissipation coefficients has a natural phenomenological 
impact and different regimes can be investigated. 

The main result of the present analysis 
is to recognize that, also in the 
presence of a fluctuating poloidal magnetic field, the 3D-turbulence picture, 
i.e. in which non-axially symmetric modes are present, has a natural branch of the non-linear dynamics along which it 
decays toward a two-dimensional profile. 
This result generalizes the study in \cite{2023PhyD..45133774M}, concerning the electrostatic turbulence only. Although the attractive nature of the axially symmetric turbulence emerges in both cases, here we found a different morphology of the 
decaying rate: 
while in the electrostatic scenario 
the growth rate depends on $n^2$ and 
a single branch is present, here 
this same quantity is associated with dissipation coefficients, but, below a 
given wavenumber, it is independent 
of the toroidal index $n$. 
Furthermore, a second branch emerged 
for wavenumbers above the threshold provided by the parameter $\bar{k}_n$ introduced in this work, and in this case a dependence on the axial number $n$ is restored.
However, it is important to remark that no peculiar limit exists to recover the pure electrostatic case from the electromagnetic analysis. It is merely this fact that clarifies the relevance of dealing with a fluctuating poloidal magnetic field, since it is able to induce, especially for high values of the $\beta$-parameter of the plasma, a specific fingerprint on the turbulent dynamics as a whole.

\section*{Appendix}\label{appendix}
Here we show that the terms containing integrals in Eqs. (\ref{gfye1}) and (\ref{gf5b}) vanish under the assumptions $\xi^0_k=\frac{\Gamma}{k^2}$ and $\xi^n_\mathbf{k}=f^n_k$. To begin, let us focus on the $n=0$ case, which we report here for the sake of clarity
\begin{equation*}
    \int_D \, dq_udq_v \, q^2
	\left(k_uq_v-k_vq_u\right) \frac{\Gamma}{|\mathbf{k}-\mathbf{q}|^2}\frac{\Gamma}{q^2}.
\end{equation*}
We express both the vectors $\mathbf{q}$ and $\mathbf{k}$ in polar coordinates, i.e.
\begin{align*}
    &q_u=r \cos{\theta} \qquad \qquad k_u=\rho \cos{\varphi}\\
    &q_v=r \sin{\theta} \qquad \qquad k_v=\rho \sin{\varphi}
\end{align*}
where $\rho$ and $\varphi$ are two constants. The integral rewrites 
\begin{equation*}
    \int_0^{k_{max}}dr \, r^4 \int_0^{2\pi}d\theta \, \sin \left ( {\theta-\varphi} \right ) \left [ \frac{1}{r^2}+\frac{1}{r^2+\rho^2-2\rho r \cos\left ( {\theta-\varphi} \right )}\right ],
\end{equation*}
where we discarded constant factors. By exploiting the periodicity of the trigonometric functions, it is immediate to verify that the expression above is identically null. We now turn our attention to the $n \neq 0$ case, in which the integral has the form
\begin{equation*}
    \int_D dq_udq_v 
	q^2\left( k_uq_v - k_vq_u\right) 
	\left[ f_{|\mathbf{k}-\mathbf{q}|}^n
	\frac{\Gamma}{q^2}+ \frac{\Gamma}{|\mathbf{k}-\mathbf{q}|^2}f_{q}^n \right]. 
\end{equation*}
By resorting, as before, to polar coordinates and expanding the Fourier mode $f^n_q$ in Taylor series, i.e.
\begin{equation*}
    f^n_q= \sum_{m=0}^\infty c^n_m \, q^m=\sum_{m=0}^\infty c^n_m \, r^m,
\end{equation*}
with $c^n_m$ a matrix of real coefficients, we observe that the integral reads
\begin{equation*}
 \int_0^{k_{max}}dr \, r^4 \int_0^{2\pi}d\theta \, \sin \left ( {\theta-\varphi} \right ) \left [ \frac{\left( r^2+\rho^2-2\rho r \cos\left ( {\theta-\varphi} \right )\right )^{\frac{m}{2}}}{r^2}+\frac{r^m}{r^2+\rho^2-2\rho r \cos\left ( {\theta-\varphi} \right )}\right ],
\end{equation*}
where $m$ is an integer non-negative number. Again, it sufficient to invoke the periodicity of the trigonometric functions to demonstrate that, also in this case, the integral turns out to be identically null.

\bibliography{biblio.bib}

\begin{thebibliography}{27}%
\makeatletter
\providecommand \@ifxundefined [1]{%
 \@ifx{#1\undefined}
}%
\providecommand \@ifnum [1]{%
 \ifnum #1\expandafter \@firstoftwo
 \else \expandafter \@secondoftwo
 \fi
}%
\providecommand \@ifx [1]{%
 \ifx #1\expandafter \@firstoftwo
 \else \expandafter \@secondoftwo
 \fi
}%
\providecommand \natexlab [1]{#1}%
\providecommand \enquote  [1]{``#1''}%
\providecommand \bibnamefont  [1]{#1}%
\providecommand \bibfnamefont [1]{#1}%
\providecommand \citenamefont [1]{#1}%
\providecommand \href@noop [0]{\@secondoftwo}%
\providecommand \href [0]{\begingroup \@sanitize@url \@href}%
\providecommand \@href[1]{\@@startlink{#1}\@@href}%
\providecommand \@@href[1]{\endgroup#1\@@endlink}%
\providecommand \@sanitize@url [0]{\catcode `\\12\catcode `\$12\catcode `\&12\catcode `\#12\catcode `\^12\catcode `\_12\catcode `\%12\relax}%
\providecommand \@@startlink[1]{}%
\providecommand \@@endlink[0]{}%
\providecommand \url  [0]{\begingroup\@sanitize@url \@url }%
\providecommand \@url [1]{\endgroup\@href {#1}{\urlprefix }}%
\providecommand \urlprefix  [0]{URL }%
\providecommand \Eprint [0]{\href }%
\providecommand \doibase [0]{http://dx.doi.org/}%
\providecommand \selectlanguage [0]{\@gobble}%
\providecommand \bibinfo  [0]{\@secondoftwo}%
\providecommand \bibfield  [0]{\@secondoftwo}%
\providecommand \translation [1]{[#1]}%
\providecommand \BibitemOpen [0]{}%
\providecommand \bibitemStop [0]{}%
\providecommand \bibitemNoStop [0]{.\EOS\space}%
\providecommand \EOS [0]{\spacefactor3000\relax}%
\providecommand \BibitemShut  [1]{\csname bibitem#1\endcsname}%
\let\auto@bib@innerbib\@empty
\bibitem [{\citenamefont {Wesson}\ and\ \citenamefont {Campbell}(2011)}]{wesson2011tokamaks}%
  \BibitemOpen
  \bibfield  {author} {\bibinfo {author} {\bibfnamefont {J.}~\bibnamefont {Wesson}}\ and\ \bibinfo {author} {\bibfnamefont {D.}~\bibnamefont {Campbell}},\ }\href {https://books.google.it/books?id=XJssMXjHUr0C} {\emph {\bibinfo {title} {Tokamaks}}},\ International Series of Monographs on Physics\ (\bibinfo  {publisher} {OUP Oxford},\ \bibinfo {year} {2011})\BibitemShut {NoStop}%
\bibitem [{\citenamefont {{Tamain}}\ \emph {et~al.}(2016)\citenamefont {{Tamain}}, \citenamefont {{Bufferand}}, \citenamefont {{Ciraolo}}, \citenamefont {{Colin}}, \citenamefont {{Galassi}}, \citenamefont {{Ghendrih}}, \citenamefont {{Schwander}},\ and\ \citenamefont {{Serre}}}]{tokam3x}%
  \BibitemOpen
  \bibfield  {author} {\bibinfo {author} {\bibfnamefont {P.}~\bibnamefont {{Tamain}}}, \bibinfo {author} {\bibfnamefont {H.}~\bibnamefont {{Bufferand}}}, \bibinfo {author} {\bibfnamefont {G.}~\bibnamefont {{Ciraolo}}}, \bibinfo {author} {\bibfnamefont {C.}~\bibnamefont {{Colin}}}, \bibinfo {author} {\bibfnamefont {D.}~\bibnamefont {{Galassi}}}, \bibinfo {author} {\bibfnamefont {P.}~\bibnamefont {{Ghendrih}}}, \bibinfo {author} {\bibfnamefont {F.}~\bibnamefont {{Schwander}}}, \ and\ \bibinfo {author} {\bibfnamefont {E.}~\bibnamefont {{Serre}}},\ }\href@noop {} {\bibfield  {journal} {\bibinfo  {journal} {Journal of Computational Physics}\ }\textbf {\bibinfo {volume} {321}},\ \bibinfo {pages} {606} (\bibinfo {year} {2016})}\BibitemShut {NoStop}%
\bibitem [{\citenamefont {{Stegmeir}}\ \emph {et~al.}(2019)\citenamefont {{Stegmeir}}, \citenamefont {{Ross}}, \citenamefont {{Body}}, \citenamefont {{Francisquez}}, \citenamefont {{Zholobenko}}, \citenamefont {{Coster}}, \citenamefont {{Maj}}, \citenamefont {{Manz}}, \citenamefont {{Jenko}}, \citenamefont {{Rogers}},\ and\ \citenamefont {{Kang}}}]{stegmeir-GRILLIX-19}%
  \BibitemOpen
  \bibfield  {author} {\bibinfo {author} {\bibfnamefont {A.}~\bibnamefont {{Stegmeir}}}, \bibinfo {author} {\bibfnamefont {A.}~\bibnamefont {{Ross}}}, \bibinfo {author} {\bibfnamefont {T.}~\bibnamefont {{Body}}}, \bibinfo {author} {\bibfnamefont {M.}~\bibnamefont {{Francisquez}}}, \bibinfo {author} {\bibfnamefont {W.}~\bibnamefont {{Zholobenko}}}, \bibinfo {author} {\bibfnamefont {D.}~\bibnamefont {{Coster}}}, \bibinfo {author} {\bibfnamefont {O.}~\bibnamefont {{Maj}}}, \bibinfo {author} {\bibfnamefont {P.}~\bibnamefont {{Manz}}}, \bibinfo {author} {\bibfnamefont {F.}~\bibnamefont {{Jenko}}}, \bibinfo {author} {\bibfnamefont {B.}~\bibnamefont {{Rogers}}}, \ and\ \bibinfo {author} {\bibfnamefont {K.}~\bibnamefont {{Kang}}},\ }\href@noop {} {\bibfield  {journal} {\bibinfo  {journal} {Phys. Plasmas}\ }\textbf {\bibinfo {volume} {26}},\ \bibinfo {pages} {052517} (\bibinfo {year} {2019})}\BibitemShut {NoStop}%
\bibitem [{\citenamefont {{Cianfrani}}\ \emph {et~al.}(2022)\citenamefont {{Cianfrani}}, \citenamefont {{Fuhr}},\ and\ \citenamefont {{Beyer}}}]{2022PhPl...29c2302C}%
  \BibitemOpen
  \bibfield  {author} {\bibinfo {author} {\bibfnamefont {F.}~\bibnamefont {{Cianfrani}}}, \bibinfo {author} {\bibfnamefont {G.}~\bibnamefont {{Fuhr}}}, \ and\ \bibinfo {author} {\bibfnamefont {P.}~\bibnamefont {{Beyer}}},\ }\href@noop {} {\bibfield  {journal} {\bibinfo  {journal} {Physics of Plasmas}\ }\textbf {\bibinfo {volume} {29}},\ \bibinfo {eid} {032302} (\bibinfo {year} {2022})}\BibitemShut {NoStop}%
\bibitem [{\citenamefont {Oliveira}\ \emph {et~al.}(2022)\citenamefont {Oliveira}, \citenamefont {Body}, \citenamefont {Galassi}, \citenamefont {Theiler}, \citenamefont {Laribi}, \citenamefont {Tamain}, \citenamefont {Stegmeir}, \citenamefont {Giacomin}, \citenamefont {Zholobenko}, \citenamefont {Ricci}, \citenamefont {Bufferand}, \citenamefont {Boedo}, \citenamefont {Ciraolo}, \citenamefont {Colandrea}, \citenamefont {Coster}, \citenamefont {de~Oliveira}, \citenamefont {Fourestey}, \citenamefont {Gorno}, \citenamefont {Imbeaux}, \citenamefont {Jenko}, \citenamefont {Naulin}, \citenamefont {Offeddu}, \citenamefont {Reimerdes}, \citenamefont {Serre}, \citenamefont {Tsui}, \citenamefont {Varini}, \citenamefont {Vianello}, \citenamefont {Wiesenberger}, \citenamefont {Wüthrich},\ and\ \citenamefont {the TCV~Team}}]{Oliveira_2022}%
  \BibitemOpen
  \bibfield  {author} {\bibinfo {author} {\bibfnamefont {D.}~\bibnamefont {Oliveira}}, \bibinfo {author} {\bibfnamefont {T.}~\bibnamefont {Body}}, \bibinfo {author} {\bibfnamefont {D.}~\bibnamefont {Galassi}}, \bibinfo {author} {\bibfnamefont {C.}~\bibnamefont {Theiler}}, \bibinfo {author} {\bibfnamefont {E.}~\bibnamefont {Laribi}}, \bibinfo {author} {\bibfnamefont {P.}~\bibnamefont {Tamain}}, \bibinfo {author} {\bibfnamefont {A.}~\bibnamefont {Stegmeir}}, \bibinfo {author} {\bibfnamefont {M.}~\bibnamefont {Giacomin}}, \bibinfo {author} {\bibfnamefont {W.}~\bibnamefont {Zholobenko}}, \bibinfo {author} {\bibfnamefont {P.}~\bibnamefont {Ricci}}, \bibinfo {author} {\bibfnamefont {H.}~\bibnamefont {Bufferand}}, \bibinfo {author} {\bibfnamefont {J.}~\bibnamefont {Boedo}}, \bibinfo {author} {\bibfnamefont {G.}~\bibnamefont {Ciraolo}}, \bibinfo {author} {\bibfnamefont {C.}~\bibnamefont {Colandrea}}, \bibinfo {author} {\bibfnamefont {D.}~\bibnamefont {Coster}}, \bibinfo {author} {\bibfnamefont {H.}~\bibnamefont
  {de~Oliveira}}, \bibinfo {author} {\bibfnamefont {G.}~\bibnamefont {Fourestey}}, \bibinfo {author} {\bibfnamefont {S.}~\bibnamefont {Gorno}}, \bibinfo {author} {\bibfnamefont {F.}~\bibnamefont {Imbeaux}}, \bibinfo {author} {\bibfnamefont {F.}~\bibnamefont {Jenko}}, \bibinfo {author} {\bibfnamefont {V.}~\bibnamefont {Naulin}}, \bibinfo {author} {\bibfnamefont {N.}~\bibnamefont {Offeddu}}, \bibinfo {author} {\bibfnamefont {H.}~\bibnamefont {Reimerdes}}, \bibinfo {author} {\bibfnamefont {E.}~\bibnamefont {Serre}}, \bibinfo {author} {\bibfnamefont {C.}~\bibnamefont {Tsui}}, \bibinfo {author} {\bibfnamefont {N.}~\bibnamefont {Varini}}, \bibinfo {author} {\bibfnamefont {N.}~\bibnamefont {Vianello}}, \bibinfo {author} {\bibfnamefont {M.}~\bibnamefont {Wiesenberger}}, \bibinfo {author} {\bibfnamefont {C.}~\bibnamefont {Wüthrich}}, \ and\ \bibinfo {author} {\bibnamefont {the TCV~Team}},\ }\href@noop {} {\bibfield  {journal} {\bibinfo  {journal} {Nuclear Fusion}\ }\textbf {\bibinfo {volume} {62}},\ \bibinfo {pages}
  {096001} (\bibinfo {year} {2022})}\BibitemShut {NoStop}%
\bibitem [{\citenamefont {Graves}\ \emph {et~al.}(2005)\citenamefont {Graves}, \citenamefont {Horacek}, \citenamefont {Pitts},\ and\ \citenamefont {Hopcraft}}]{Graves_2005}%
  \BibitemOpen
  \bibfield  {author} {\bibinfo {author} {\bibfnamefont {J.~P.}\ \bibnamefont {Graves}}, \bibinfo {author} {\bibfnamefont {J.}~\bibnamefont {Horacek}}, \bibinfo {author} {\bibfnamefont {R.~A.}\ \bibnamefont {Pitts}}, \ and\ \bibinfo {author} {\bibfnamefont {K.~I.}\ \bibnamefont {Hopcraft}},\ }\href@noop {} {\bibfield  {journal} {\bibinfo  {journal} {Plasma Physics and Controlled Fusion}\ }\textbf {\bibinfo {volume} {47}},\ \bibinfo {pages} {L1} (\bibinfo {year} {2005})}\BibitemShut {NoStop}%
\bibitem [{\citenamefont {Zweben}\ \emph {et~al.}(2007)\citenamefont {Zweben}, \citenamefont {Boedo}, \citenamefont {Grulke}, \citenamefont {Hidalgo}, \citenamefont {LaBombard}, \citenamefont {Maqueda}, \citenamefont {Scarin},\ and\ \citenamefont {Terry}}]{Zweben_2007}%
  \BibitemOpen
  \bibfield  {author} {\bibinfo {author} {\bibfnamefont {S.~J.}\ \bibnamefont {Zweben}}, \bibinfo {author} {\bibfnamefont {J.~A.}\ \bibnamefont {Boedo}}, \bibinfo {author} {\bibfnamefont {O.}~\bibnamefont {Grulke}}, \bibinfo {author} {\bibfnamefont {C.}~\bibnamefont {Hidalgo}}, \bibinfo {author} {\bibfnamefont {B.}~\bibnamefont {LaBombard}}, \bibinfo {author} {\bibfnamefont {R.~J.}\ \bibnamefont {Maqueda}}, \bibinfo {author} {\bibfnamefont {P.}~\bibnamefont {Scarin}}, \ and\ \bibinfo {author} {\bibfnamefont {J.~L.}\ \bibnamefont {Terry}},\ }\href@noop {} {\bibfield  {journal} {\bibinfo  {journal} {Plasma Physics and Controlled Fusion}\ }\textbf {\bibinfo {volume} {49}},\ \bibinfo {pages} {S1} (\bibinfo {year} {2007})}\BibitemShut {NoStop}%
\bibitem [{\citenamefont {{Scott}}(2002)}]{scott02}%
  \BibitemOpen
  \bibfield  {author} {\bibinfo {author} {\bibfnamefont {B.}~\bibnamefont {{Scott}}},\ }\href@noop {} {\bibfield  {journal} {\bibinfo  {journal} {New J. Phys.}\ }\textbf {\bibinfo {volume} {52}},\ \bibinfo {pages} {352} (\bibinfo {year} {2002})}\BibitemShut {NoStop}%
\bibitem [{\citenamefont {Scott}(2007)}]{Scott_2007}%
  \BibitemOpen
  \bibfield  {author} {\bibinfo {author} {\bibfnamefont {B.~D.}\ \bibnamefont {Scott}},\ }\href@noop {} {\bibfield  {journal} {\bibinfo  {journal} {Plasma Physics and Controlled Fusion}\ }\textbf {\bibinfo {volume} {49}},\ \bibinfo {pages} {S25} (\bibinfo {year} {2007})}\BibitemShut {NoStop}%
\bibitem [{\citenamefont {{Hasegawa}}\ and\ \citenamefont {{Wakatani}}(1983)}]{hase-waka83}%
  \BibitemOpen
  \bibfield  {author} {\bibinfo {author} {\bibfnamefont {A.}~\bibnamefont {{Hasegawa}}}\ and\ \bibinfo {author} {\bibfnamefont {M.}~\bibnamefont {{Wakatani}}},\ }\href@noop {} {\bibfield  {journal} {\bibinfo  {journal} {Phys. Rev. Lett.}\ }\textbf {\bibinfo {volume} {50}},\ \bibinfo {pages} {682} (\bibinfo {year} {1983})}\BibitemShut {NoStop}%
\bibitem [{\citenamefont {{Hasegawa}}\ and\ \citenamefont {{Wakatani}}(1987)}]{hase-waka87}%
  \BibitemOpen
  \bibfield  {author} {\bibinfo {author} {\bibfnamefont {A.}~\bibnamefont {{Hasegawa}}}\ and\ \bibinfo {author} {\bibfnamefont {M.}~\bibnamefont {{Wakatani}}},\ }\href@noop {} {\bibfield  {journal} {\bibinfo  {journal} {Phys. Rev. Lett.}\ }\textbf {\bibinfo {volume} {57}},\ \bibinfo {pages} {1581} (\bibinfo {year} {1987})}\BibitemShut {NoStop}%
\bibitem [{\citenamefont {{Hasegawa}}\ and\ \citenamefont {{Mima}}(2018)}]{hase-mima18}%
  \BibitemOpen
  \bibfield  {author} {\bibinfo {author} {\bibfnamefont {A.}~\bibnamefont {{Hasegawa}}}\ and\ \bibinfo {author} {\bibfnamefont {K.}~\bibnamefont {{Mima}}},\ }\href@noop {} {\bibfield  {journal} {\bibinfo  {journal} {Eur. Phys. J. H}\ }\textbf {\bibinfo {volume} {43}},\ \bibinfo {pages} {499} (\bibinfo {year} {2018})}\BibitemShut {NoStop}%
\bibitem [{\citenamefont {Diamond}\ \emph {et~al.}(2011)\citenamefont {Diamond}, \citenamefont {Hasegawa},\ and\ \citenamefont {Mima}}]{Diamond_2011}%
  \BibitemOpen
  \bibfield  {author} {\bibinfo {author} {\bibfnamefont {P.~H.}\ \bibnamefont {Diamond}}, \bibinfo {author} {\bibfnamefont {A.}~\bibnamefont {Hasegawa}}, \ and\ \bibinfo {author} {\bibfnamefont {K.}~\bibnamefont {Mima}},\ }\href@noop {} {\bibfield  {journal} {\bibinfo  {journal} {Plasma Physics and Controlled Fusion}\ }\textbf {\bibinfo {volume} {53}},\ \bibinfo {pages} {124001} (\bibinfo {year} {2011})}\BibitemShut {NoStop}%
\bibitem [{\citenamefont {Kraichnan}\ and\ \citenamefont {Montgomery}(1980)}]{R_H_Kraichnan_1980}%
  \BibitemOpen
  \bibfield  {author} {\bibinfo {author} {\bibfnamefont {R.~H.}\ \bibnamefont {Kraichnan}}\ and\ \bibinfo {author} {\bibfnamefont {D.}~\bibnamefont {Montgomery}},\ }\href@noop {} {\bibfield  {journal} {\bibinfo  {journal} {Reports on Progress in Physics}\ }\textbf {\bibinfo {volume} {43}},\ \bibinfo {pages} {547} (\bibinfo {year} {1980})}\BibitemShut {NoStop}%
\bibitem [{\citenamefont {Boffetta}\ and\ \citenamefont {Ecke}(2012)}]{annurev:/content/journals/10.1146/annurev-fluid-120710-101240}%
  \BibitemOpen
  \bibfield  {author} {\bibinfo {author} {\bibfnamefont {G.}~\bibnamefont {Boffetta}}\ and\ \bibinfo {author} {\bibfnamefont {R.~E.}\ \bibnamefont {Ecke}},\ }\href@noop {} {\bibfield  {journal} {\bibinfo  {journal} {Annual Review of Fluid Mechanics}\ }\textbf {\bibinfo {volume} {44}},\ \bibinfo {pages} {427} (\bibinfo {year} {2012})}\BibitemShut {NoStop}%
\bibitem [{\citenamefont {Kraichnan}(1967)}]{10.1063/1.1762301}%
  \BibitemOpen
  \bibfield  {author} {\bibinfo {author} {\bibfnamefont {R.~H.}\ \bibnamefont {Kraichnan}},\ }\href@noop {} {\bibfield  {journal} {\bibinfo  {journal} {The Physics of Fluids}\ }\textbf {\bibinfo {volume} {10}},\ \bibinfo {pages} {1417} (\bibinfo {year} {1967})}\BibitemShut {NoStop}%
\bibitem [{\citenamefont {Kraichnan}(1975)}]{Kraichnan_1975}%
  \BibitemOpen
  \bibfield  {author} {\bibinfo {author} {\bibfnamefont {R.~H.}\ \bibnamefont {Kraichnan}},\ }\href@noop {} {\bibfield  {journal} {\bibinfo  {journal} {Journal of Fluid Mechanics}\ }\textbf {\bibinfo {volume} {67}},\ \bibinfo {pages} {155–175} (\bibinfo {year} {1975})}\BibitemShut {NoStop}%
\bibitem [{\citenamefont {{Biskamp}}\ and\ \citenamefont {{Zeiler}}(1995)}]{biskamp95}%
  \BibitemOpen
  \bibfield  {author} {\bibinfo {author} {\bibfnamefont {D.}~\bibnamefont {{Biskamp}}}\ and\ \bibinfo {author} {\bibfnamefont {A.}~\bibnamefont {{Zeiler}}},\ }\href@noop {} {\bibfield  {journal} {\bibinfo  {journal} {Phys. Rev. Lett.}\ }\textbf {\bibinfo {volume} {74}},\ \bibinfo {pages} {706} (\bibinfo {year} {1995})}\BibitemShut {NoStop}%
\bibitem [{\citenamefont {{Cianfrani}}\ and\ \citenamefont {{Montani}}(2024)}]{2024arXiv240509837C}%
  \BibitemOpen
  \bibfield  {author} {\bibinfo {author} {\bibfnamefont {F.}~\bibnamefont {{Cianfrani}}}\ and\ \bibinfo {author} {\bibfnamefont {G.}~\bibnamefont {{Montani}}},\ }\href@noop {} {\bibfield  {journal} {\bibinfo  {journal} {arXiv e-prints}\ ,\ \bibinfo {eid} {arXiv:2405.09837}} (\bibinfo {year} {2024})},\ \Eprint {http://arxiv.org/abs/2405.09837} {arXiv:2405.09837 [physics.plasm-ph]} \BibitemShut {NoStop}%
\bibitem [{\citenamefont {{Montani}}\ and\ \citenamefont {{Carlevaro}}(2023)}]{2023PhyD..45133774M}%
  \BibitemOpen
  \bibfield  {author} {\bibinfo {author} {\bibfnamefont {G.}~\bibnamefont {{Montani}}}\ and\ \bibinfo {author} {\bibfnamefont {N.}~\bibnamefont {{Carlevaro}}},\ }\href@noop {} {\bibfield  {journal} {\bibinfo  {journal} {Physica D Nonlinear Phenomena}\ }\textbf {\bibinfo {volume} {451}},\ \bibinfo {eid} {133774} (\bibinfo {year} {2023})}\BibitemShut {NoStop}%
\bibitem [{\citenamefont {{Montani}}\ \emph {et~al.}(2022)\citenamefont {{Montani}}, \citenamefont {{Carlevaro}},\ and\ \citenamefont {{Tirozzi}}}]{2022Fluid...7..157M}%
  \BibitemOpen
  \bibfield  {author} {\bibinfo {author} {\bibfnamefont {G.}~\bibnamefont {{Montani}}}, \bibinfo {author} {\bibfnamefont {N.}~\bibnamefont {{Carlevaro}}}, \ and\ \bibinfo {author} {\bibfnamefont {B.}~\bibnamefont {{Tirozzi}}},\ }\href@noop {} {\bibfield  {journal} {\bibinfo  {journal} {Fluids}\ }\textbf {\bibinfo {volume} {7}},\ \bibinfo {pages} {157} (\bibinfo {year} {2022})}\BibitemShut {NoStop}%
\bibitem [{\citenamefont {{Carlevaro}}\ \emph {et~al.}(2023)\citenamefont {{Carlevaro}}, \citenamefont {{Montani}},\ and\ \citenamefont {{Moretti}}}]{2023Symm...15.1745C}%
  \BibitemOpen
  \bibfield  {author} {\bibinfo {author} {\bibfnamefont {N.}~\bibnamefont {{Carlevaro}}}, \bibinfo {author} {\bibfnamefont {G.}~\bibnamefont {{Montani}}}, \ and\ \bibinfo {author} {\bibfnamefont {F.}~\bibnamefont {{Moretti}}},\ }\href@noop {} {\bibfield  {journal} {\bibinfo  {journal} {Symmetry}\ }\textbf {\bibinfo {volume} {15}},\ \bibinfo {eid} {1745} (\bibinfo {year} {2023})}\BibitemShut {NoStop}%
\bibitem [{\citenamefont {{Albanese}}\ \emph {et~al.}(2019)\citenamefont {{Albanese}}, \citenamefont {{Crisanti}}, \citenamefont {{Martin}}, \citenamefont {{Pizzuto}}, \citenamefont {{Mazzitelli}}, \citenamefont {{Tuccillo}}, \citenamefont {{Ambrosino}}, \citenamefont {{Appi}}, \citenamefont {{Di Zenobio}}, \citenamefont {{Frattolillo}}, \citenamefont {{Granucci}}, \citenamefont {{Innocente}}, \citenamefont {{Lampasi}}, \citenamefont {{Martone}}, \citenamefont {{Polli}}, \citenamefont {{Ramogida}}, \citenamefont {{Rossi}}, \citenamefont {{Sandri}}, \citenamefont {{Valisa}}, \citenamefont {{Villari}},\ and\ \citenamefont {{Vitale}}}]{dtt19}%
  \BibitemOpen
  \bibfield  {author} {\bibinfo {author} {\bibfnamefont {R.}~\bibnamefont {{Albanese}}}, \bibinfo {author} {\bibfnamefont {F.}~\bibnamefont {{Crisanti}}}, \bibinfo {author} {\bibfnamefont {P.}~\bibnamefont {{Martin}}}, \bibinfo {author} {\bibfnamefont {A.}~\bibnamefont {{Pizzuto}}}, \bibinfo {author} {\bibfnamefont {G.}~\bibnamefont {{Mazzitelli}}}, \bibinfo {author} {\bibfnamefont {A.}~\bibnamefont {{Tuccillo}}}, \bibinfo {author} {\bibfnamefont {R.}~\bibnamefont {{Ambrosino}}}, \bibinfo {author} {\bibfnamefont {G.}~\bibnamefont {{Appi}}, \bibfnamefont {A.~{Di Gironimo}}}, \bibinfo {author} {\bibfnamefont {A.}~\bibnamefont {{Di Zenobio}}}, \bibinfo {author} {\bibfnamefont {A.}~\bibnamefont {{Frattolillo}}}, \bibinfo {author} {\bibfnamefont {G.}~\bibnamefont {{Granucci}}}, \bibinfo {author} {\bibfnamefont {P.}~\bibnamefont {{Innocente}}}, \bibinfo {author} {\bibfnamefont {A.}~\bibnamefont {{Lampasi}}}, \bibinfo {author} {\bibfnamefont {R.}~\bibnamefont {{Martone}}}, \bibinfo {author} {\bibfnamefont
  {G.}~\bibnamefont {{Polli}}}, \bibinfo {author} {\bibfnamefont {G.}~\bibnamefont {{Ramogida}}}, \bibinfo {author} {\bibfnamefont {P.}~\bibnamefont {{Rossi}}}, \bibinfo {author} {\bibfnamefont {S.}~\bibnamefont {{Sandri}}}, \bibinfo {author} {\bibfnamefont {M.}~\bibnamefont {{Valisa}}}, \bibinfo {author} {\bibfnamefont {R.}~\bibnamefont {{Villari}}}, \ and\ \bibinfo {author} {\bibfnamefont {V.}~\bibnamefont {{Vitale}}},\ }\href@noop {} {\bibfield  {journal} {\bibinfo  {journal} {Fus. Eng. Des.}\ }\textbf {\bibinfo {volume} {146}},\ \bibinfo {pages} {194} (\bibinfo {year} {2019})}\BibitemShut {NoStop}%
\bibitem [{\citenamefont {{Ambrosino}}(2021)}]{DTT_2021}%
  \BibitemOpen
  \bibfield  {author} {\bibinfo {author} {\bibfnamefont {R.}~\bibnamefont {{Ambrosino}}},\ }\href@noop {} {\bibfield  {journal} {\bibinfo  {journal} {Fus. Eng. Des.}\ }\textbf {\bibinfo {volume} {167}},\ \bibinfo {pages} {112330} (\bibinfo {year} {2021})}\BibitemShut {NoStop}%
\bibitem [{\citenamefont {{Seyler}}\ \emph {et~al.}(1975)\citenamefont {{Seyler}}, \citenamefont {{Salu}}, \citenamefont {{Montgomery}},\ and\ \citenamefont {{Knorr}}}]{seyler75}%
  \BibitemOpen
  \bibfield  {author} {\bibinfo {author} {\bibfnamefont {C.}~\bibnamefont {{Seyler}}}, \bibinfo {author} {\bibfnamefont {Y.}~\bibnamefont {{Salu}}}, \bibinfo {author} {\bibfnamefont {D.}~\bibnamefont {{Montgomery}}}, \ and\ \bibinfo {author} {\bibfnamefont {G.}~\bibnamefont {{Knorr}}},\ }\href@noop {} {\bibfield  {journal} {\bibinfo  {journal} {Phys. Fluids}\ }\textbf {\bibinfo {volume} {18}},\ \bibinfo {pages} {803} (\bibinfo {year} {1975})}\BibitemShut {NoStop}%
\bibitem [{\citenamefont {Ryutov}\ and\ \citenamefont {Soukhanovskii}(2015)}]{10.1063/1.4935115}%
  \BibitemOpen
  \bibfield  {author} {\bibinfo {author} {\bibfnamefont {D.~D.}\ \bibnamefont {Ryutov}}\ and\ \bibinfo {author} {\bibfnamefont {V.~A.}\ \bibnamefont {Soukhanovskii}},\ }\href@noop {} {\bibfield  {journal} {\bibinfo  {journal} {Physics of Plasmas}\ }\textbf {\bibinfo {volume} {22}},\ \bibinfo {pages} {110901} (\bibinfo {year} {2015})}\BibitemShut {NoStop}%
\bibitem [{\citenamefont {Sonnino}\ \emph {et~al.}(2014)\citenamefont {Sonnino}, \citenamefont {Cardinali}, \citenamefont {Sonnino}, \citenamefont {Nardone}, \citenamefont {Steinbrecher},\ and\ \citenamefont {Zonca}}]{10.1063/1.4867656}%
  \BibitemOpen
  \bibfield  {author} {\bibinfo {author} {\bibfnamefont {G.}~\bibnamefont {Sonnino}}, \bibinfo {author} {\bibfnamefont {A.}~\bibnamefont {Cardinali}}, \bibinfo {author} {\bibfnamefont {A.}~\bibnamefont {Sonnino}}, \bibinfo {author} {\bibfnamefont {P.}~\bibnamefont {Nardone}}, \bibinfo {author} {\bibfnamefont {G.}~\bibnamefont {Steinbrecher}}, \ and\ \bibinfo {author} {\bibfnamefont {F.}~\bibnamefont {Zonca}},\ }\href@noop {} {\bibfield  {journal} {\bibinfo  {journal} {Chaos: An Interdisciplinary Journal of Nonlinear Science}\ }\textbf {\bibinfo {volume} {24}},\ \bibinfo {pages} {013129} (\bibinfo {year} {2014})}\BibitemShut {NoStop}%
\end{thebibliography}%
\end{document}